\documentclass{hotnets17}
\usepackage{enumitem}
\usepackage{times}
\pdfoutput=1 
\usepackage{xspace}
\usepackage{graphicx}
\usepackage{multirow}
\usepackage{array}
\usepackage{color}
\usepackage{flushend}
\usepackage{fancyvrb}
\usepackage{times}
\usepackage{amsfonts}
\usepackage{amsmath}
\usepackage{amsmath,amssymb}
\usepackage{soul}
\DeclareMathOperator{\E}{\mathbb{E}}
\usepackage{algorithm}
\usepackage[noend]{algpseudocode}

\newcommand{\parab}[1]{\vspace*{0.07in}\noindent\textbf{#1}}

\newcommand{\mycomment}[1]{}
\newcommand{\ignore}[1]{}
\newcommand{\url}[1]{\texttt{\small{#1}}}

\newcommand{\toolshort}{DeepConf\xspace}
\newcommand{\sysname}{DeepConf\xspace}
\newcommand{\sysnameagent}{\sysname-agent\xspace}
\newcommand{\sysnameagents}{\sysname-agents\xspace}

\makeatletter
\def\BState{\State\hskip-\ALG@thistlm}
\makeatother

\usepackage{bbm, bm}
\numberwithin{equation}{section}

\newlength{\dhatheight}

\renewcommand{\P}{\mathbb{P}}

\begin{document}

\title{\toolshort: Automating Data Center Network Topologies Management with Machine Learning}
 \author{Christopher Streiffer$^*$, Huan Chen$^{* \dagger}$, Theophilus Benson$^+$, Asim Kadav$^\ddagger$\\
 Duke University$^*$ \quad UESTC, China$^\dagger$ \quad Brown University$^+$   \quad NEC Labs$^\ddagger$ }
%\author{Paper \#94}
%\subtitle{}
\date{}
\maketitle
\thispagestyle{empty}

%------------------------------------------------------------------------------
%                                Main Contents.
%------------------------------------------------------------------------------
\begin{abstract}
In recent years, many techniques have been developed to improve the performance and efficiency of data center networks. While these techniques provide high accuracy, they are often designed using heuristics that leverage domain-specific properties of the workload or hardware.

In this vision paper, we argue that many data center networking techniques, e.g., routing, topology augmentation, energy savings, with diverse goals actually share design and architectural similarity. We present a design for developing general intermediate representations of network topologies using deep learning that is amenable to solving classes of data center problems. We develop a framework, \sysname, that simplifies the processing of configuring and training deep learning agents that use the intermediate representation to learns different tasks. To illustrate the strength of our approach, we configured, implemented, and evaluated a \sysnameagent that tackles the data center topology augmentation problem. Our initial results are promising --- \sysname performs comparably to the optimal.

\end{abstract}

%!TEX root = ../main.tex
\section{Introduction}
Data center networks (DCN) are a crucial and important part of the Internet's ecosystem. The performance of these DCNs impact a wide variety of services from web browsing and videos to Internet of Things.  The poor performance of these DCNs can result in as much as \$4 million in lost revenue~\cite{gigaspaces}. 

Motivated by the importance of these networks, the networking community has explored techniques for improving and managing the performance of the data center network topology by: (1) designing better routing or traffic engineering algorithms~\cite{hedera:nsdi10, microte:conext11,drill:hotnets15,flowcomb:hotcloud13}, (2) improving performance of a fixed topology by adding a limited number of flexible links~\cite{wang:sigcomm10,helios:sigcomm10,hamedazimi:sigcomm14,halperin:sigcomm11}, and (3) removing corrupted and underutilized links from the topology to save energy and improve performance~\cite{elastictree:nsdi10,Greenberg:2008:CCR:1496091.1496103, corropt:sigcomm17}.

Regardless of the approach, these topology-oriented techniques have three things in common: (1) Each is formalized as an optimization problem.
(2) Due to the impracticality of scalably solving these optimizations, greedy heuristics are employed to create approximate solutions. (3) Each heuristic is intricately tied to the application patterns and does not generalize across novel patterns. 
Existing domain-specific heuristics provide suboptimal performance and are often limited to specific scenarios. Thus as a community, we are forced to revisit and redesign these heuristics when the application pattern or network details changes -- even a minor change. For example, while c-through~\cite{wang:sigcomm10} and FireFly~\cite{hamedazimi:sigcomm14} solve broadly identical problems, they leverage different heuristics to account for low-level differences.

In this paper, we articulate our vision for replacing domain-specific rule-based heuristics for topology management with a more general machine learning-based (ML) model that quickly learns optimal solutions to a class of problems while adapting to changes in the application patterns, network dynamics, and low-level network details. 
Unlike recent attempts that employ ML to learn point solutions, e.g., cluster scheduling~\cite{mao2016resource} or routing~\cite{boyan1994packet}, in this paper, 
we present a general framework, called \sysname, that simplifies the process of designing ML models for a broad range of DCN topology problems and eliminates the challenges associated with efficiently training new models.

The key challenges in designing \sysname are: (1) tackling the dichotomy that exists between deep learning's requirements for large amounts of supervised data and the unavailability of these required datasets, and (2) designing a general, but highly accurate, deep learning model that efficiently generalizes to learning a broad array of data center problems ranging from topology management and routing to energy savings.

The key insight underlying \sysname is that intermediate features generated from the parallel convolutional layers using network data, e.g., traffic matrix, allows us to generate an intermediate representation of the network's state that enables learning a broad range of data center problems.  Moreover, while labeled production data crucial for machine learning is unavailable, empirical studies~\cite{morpheus} show that modern data center traffic is higly predictable and thus amenable to offline learning with network simulators and historical traces.

\sysname builds on this insight by using reinforcement learning (RL), an unsupervised technique, that learns through experience and makes no assumptions on how the network works. Rather, they are trained through the use of a reward signal which ``guides'' them towards an optimal solution and thus do not require real world data, and, instead, they can be trained using simulators. 

The \sysname framework provides a predefined RL model with the intermediate representation, a specific design for configuring this model to address different problems, an optimized simulator to enable efficient learning, and an SDN-based platform for capturing network data and reconfiguring the network.

In this paper, we make the following contributions:
\begin{itemize}[leftmargin=*]
	\item We present a novel RL-based SDN architecture for developing and training deep ML models for a broad range of DCN tasks.
	\item We design a novel input feature extraction for DCNs for developing different ML models over this intermediate representation of network state.
	\item We implemented a \sysnameagent tackling the topology augmentation problem and evaluated it on representative topologies~\cite{vl2, fattree} and traces~\cite{fb:trace}, showing that our approach performs comparable to optimal.
	
\end{itemize}

\section{Related Work}
\label{sec:related}
%related here.
% Recent RL work
Our work is motivated by the recent success of applying machine learning and RL algorithms to computer games and robotic planning~\cite{mnih2013playing, silver2016mastering, usunier2016episodic}. The most closely related work~\cite{boyan1994packet} applies RL to packet routing. Unlike~\cite{boyan1994packet}, \sysname tackles the topology augmentation problem and explores the use of deep networks as function approximators for RL.
Existing applications of machine learning to data centers focus on improving cluster scheduling~\cite{mao2016resource} and more recently by Google to optimize Power Usage Effectiveness(PUE)~\cite{google:deepmind}. 
In this vision paper, we take a different stance and focus on identifying a class of equivalent data center management operations, namely topology management and configuration, that are amenable to a common machine learning approach and design a modular system that enables different agents to interoperate over a network.

%
% There has been resurgence in reinforcement learning algorithms
% using neural networks as function approximators. These algorithms
% have been used across a broad spectrum of domains from computer games and robotic planning to
% and control and other applications~\cite{mnih2013playing, silver2016mastering, usunier2016episodic}.
%
% % ML/RL uses in systems (perhaps fold in intro/background)
% Machine learning has been used previously to address datacenter
% configurations. Reinforcement learning has been used for
% network packet routing~\cite{boyan1994packet}. However, this
% work does not use deep networks as function approximators.
% Deepmind demonstrated using machine learning to optimize PUE
% (Power Usage Effectiveness) which measure power utilization over effective
% IT usage. They find that using deep learning networks
% to predict temperature and pressure uses the lowest possible PUE and
% is effective. RL has also been used in the context of resource management
% specifically for resource packing~\cite{mao2016resource} using the standard
% REINFORCE algorithm. \toolshort\ uses a two-stream model architecture with
% asynchronous learning to speed up evaluations and learn appropriate network actions.

%!TEX root = ../main.tex
\section{Background}
\label{sec:back}
This section provides an overview of data center networking challenges and solutions and provides background on our reinforcement learning methods.

\subsection{Data Center Networks}
Data center networks introduce a range of challenges from topology design and routing algorithms to VM placement and energy saving techniques.  

Data centers support a large variety of workloads and applications with time-varying bandwidth requirements. This variance in bandwidth requirements leads to hotspots at varying locations and at different points in time. To support these arbitrary bandwidth requirements, data center operators can employ non-blocking topologies; however, non-blocking topologies are prohibitively costly. Instead, these operators employ a range of techniques ranging from hybrid architectures~\cite{wang:sigcomm10,helios:sigcomm10,hamedazimi:sigcomm14,halperin:sigcomm11}, traffic engineering algorithms~\cite{hedera:nsdi10, microte:conext11,drill:hotnets15,flowcomb:hotcloud13}, and energy saving techniques~\cite{elastictree:nsdi10,Nedevschi:2008:RNE:1387589.1387612}.  Below, we describe these techniques and illustrate common designs.
%\begin{itemize} 

\begin{figure}[!tb]
\centering
\includegraphics[width=0.6\columnwidth]{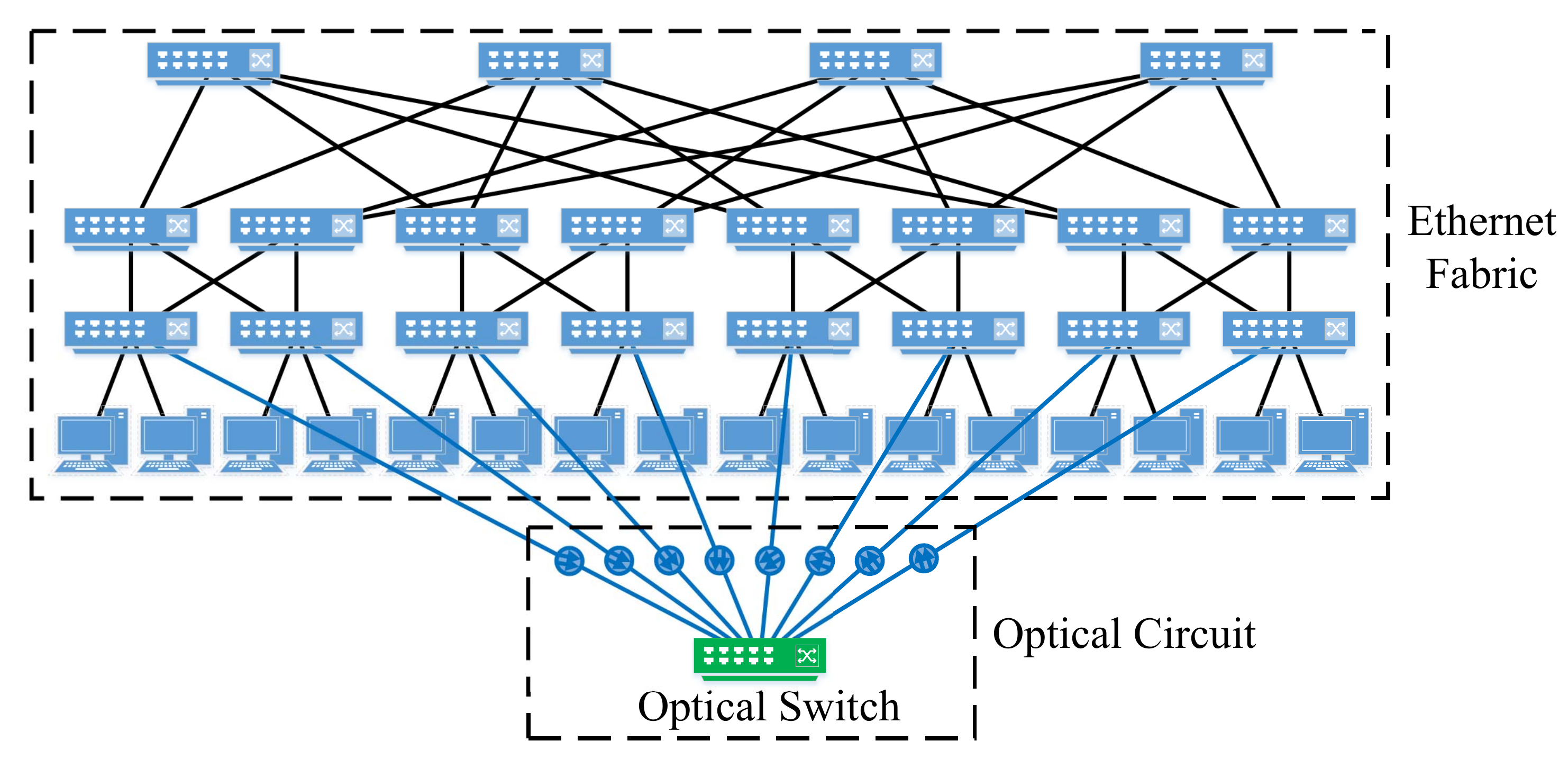}
\vspace{-0.2in}
\caption{\small \bf One optical switch with k=4 Fat Tree Topology.}
\vspace{-0.25in}
%The optic switch (center) connects the edge switches together. The optical switch allows for up to $n$ connections to be made between edge switches at a time.}
\label{fig:opswitch}
\end{figure}

\parab{Augmented Architectures:} This class of approaches build on the intuition that at any given point in time, there are only a small number of hotspots (or congested links). Thus, there is no need to build an expensive topology that supports full bisection (eliminating all potential points of congestion). Instead, the existing topology can be augmented with a small number of links which can be added ondemand and moved to the location of the hotspot or congested links. 
%Additionally, over time these links can be moved around to different locations as the hotspots move. 
These approaches augment the data center's ethernet network with a small number of optical~\cite{wang:sigcomm10,helios:sigcomm10}, wireless~\cite{halperin:sigcomm11}, or free optics~\cite{hamedazimi:sigcomm14}.~\footnote{The number of augmented links is significantly smaller than the number of data center's permanent links.} 
For example, Figure~\ref{fig:opswitch} shows a traditional Fat-Tree topology augmented with an optical switch -- as was proposed by Helios~\cite{helios:sigcomm10}. 

%At a high-level, these approach rapidly shift these augmented links around the data center to locations with hotspots. These augmented links often take a while to 
%To adapt to these time-varying characteristics. Several have proposed augmenting the data center topology with additional flexible link, e.g., Optical links~\cite{}, wireless links~\cite{}, and free-optics~\cite{}. In all proposals, there are a limited number of flexible links and the number of flexible links is much smaller than the number of fixed links within the topology. Additionally, these moving these flexible links takes a significant amount of time, so care must be taken to ensure efficient placement. 
These proposals argue for monitoring the traffic, using an Integer Linear Program (ILP) or heuristic to detect the hotspots and place the flexible links at the location with these hotspots. Unfortunately, moving these flexible links incurs a large switching time during which the links are not operational. These intelligent and efficient algorithms are developed to effectively detect hotspots and efficiently place links.

\parab{Traffic Engineering:} Orthogonal approaches~\cite{hedera:nsdi10, microte:conext11,drill:hotnets15,flowcomb:hotcloud13} focus on routing. Instead of changing the topologies, these approaches change the mapping of flows to paths within a fixed topology.
%Data centers contain a mixture of small flows and large flows. These small flows often require low latency but their latency is impacted by large flows. To address these issues, many techniques~\cite{} have been proposed to improve latency by detecting and migrating elephant flows.
These proposals, also, argue for monitoring traffic and detecting hotspots. Instead of changing topologies, these techniques move a subset of flows from congested links to un-congested links.

\parab{Energy Savings} Data centers are notorious for their energy usage~\cite{Greenberg:2008:CCR:1496091.1496103}. To address this, researchers have proposed techniques to improve energy efficiency by detecting periods of low utilization and selectively turning off links~\cite{elastictree:nsdi10,Nedevschi:2008:RNE:1387589.1387612}. These proposals argue for monitoring traffic, detecting low utilization, and powering-down links in portions of the data center with low utilizations. A key challenge with these techniques is to turn on the powered-down links before demands rise. 
%\end{itemize}
\newline

Taking a step back, these techniques roughly follow the same design and operate in three steps (1) gather network traffic matrix, (2) run an ILP to predict heavy (or low) usage, and (3) perform a specific action on a subset of the network. The actions range from augmenting flexible links, turning off links, or moving traffic. In all situations, the ILP does not scale to a large network and a domain-specific heuristic is often used in its place.

% These set of tasks are ideal for a form of deep learning called deep reinforcement learning -- where in an algorithm learns through experience the set of actions to perform under a variety of conditions. Unlike supervised learning techniques which require a significantly large corpus of labeled-data which is prohibitively expensive, unsupervised learning techniques, e.g., DeepRL, learns from unlabeled-data and merely requires data sets and a simulation.

\subsection{Reinforcement Learning}

Reinforcement learning (RL) algorithms learn through experience with a goal towards maximizing rewards. Unlike supervised learning where algorithms train over labels, RL algorithms learn by interacting with an environment such as a game or a network simulator. 

In traditional RL, an agent interacts with an environment over a number of discrete time steps. Hence, at each time step $t$, the agent in a {\em world} observes a state $s_t$ in order to select an action $a_t$ from a possible action set $A$.
% However, unlike in a simulator, the agent is not aware of how the world works i.e. the specific reward function. Instead, 
The agent is guided by a policy, $\pi$, which is a function that maps state $s_t$ to actions $a_t$. The agent receives a reward $r_t$ for each action and transitions to the next state $s_{t+1}$. The goal of the agent is maximizing the total reward. This process continues until the agent reaches a final state or time limit, after which the environment is reset and a new training episode is played. After a number of training episodes, the agent learns to pick actions that maximize the rewards and can learn to handle unexpected states. RL is effective and has been successfully used to model robotics, game bots, etc.

The goal of commonly used policy-based RL is to find a policy, $\pi$, that maximizes the cumulative reward and converges to a theoretical optimal policy. In deep policy-based methods, a neural network computes a policy distribution $\pi(a_t|s_t;\theta)$, where $\theta$ represents the set of parameters of the function. Deep networks as function approximators is a recent development and other learning methods can be used. We now describe the REINFORCE and actor-critic policy methods which represent different methods to score the policy $J(\theta)$. REINFORCE methods \cite{williams1992simple} use gradient ascent on $\E[R_t]$, where $R_t=\sum_{i=0}^{\infty}\gamma^ir_{t+i}$ is the  accumulated reward starting from time step $t$ and discounted at each step by $\gamma\in (0, 1]$, the discount factor. The REINFORCE method, which is the Monte-Carlo method, updates $\theta$ using the gradient $\nabla_\theta\log\pi(a_t|s_t;\theta)R_t$, which is an unbiased estimator of $\nabla_\theta\E[R_t]$. %The variance of the estimator is reduced by subtracting a learned \textit{baseline} (a function of the state $b_t(s_t)$) and using the gradient $\nabla_\theta\log\pi(a_t|s_t;\theta)\big(R_t~-~b_t(s_t)\big)$ instead. T
The value function is computed as $V^\pi(s_t) = \E[{R_t|s_t}]$ which is the expected return for following the policy $\pi$ in state $s_t$. This method provides actions with high returns but suffers from high-variance of gradient estimates.

{\bf Asynchronous Advantage Actor Critic (A3C)}: A3C~\cite{mnih2016asynchronous} improves REINFORCE performance by operating asynchronously and by using a deep network to approximate the policy and value faction. A3C uses the actor-critic method which additionally computes a critic function that approximates the value function. A3C, as used by us, uses a network with two convolutional layers followed by a fully connected layer. Each hidden layer is followed by a nonlinearity function (ReLU). A softmax layer which approximates the policy function and a linear layer to output an estimate of the value function $V \left(s_t; \theta\right)$ together constitute the output of this network. Asynchronous gradient descent using multiple agents is used to train the network and this improves the training speed. A central server (similar to a parameter server) coordinates the parallel agents -- each agent calculates the gradients and sends the updates to the server after a fixed number of steps, or when a final state is reached. Furthermore, following each update, the central server propagates new weights to the agents to achieve a consensus on the policy values. There is a cost function with each deep network (policy and value). Using two loss functions has found to improve convergence and produce better-regularized models.
The policy cost function is given as:

\vspace{-0.2in}
\begin{equation}
f_{\pi}\left( \theta \right)
 = \log\pi\left(a_t | s_t; \theta \right)
\left(R_t - V\left(s_t; \theta_t \right) \right)
+ \beta H \left(\pi \left(s_t; \theta \right)\right)
\label{eq:costPi}
\end{equation}
where $\theta_t$ represents the values of the parameters $\theta$ at time $t$, $R_t = \sum_{i=0}^{k-1}{\gamma^i r_{t+i} + \gamma^k V\left(s_{t+k}; \theta_t \right)}$ is the estimated discounted reward.% in the time interval from $t$ to $t+k$ and $k$ is upper-bounded by $t_{max}$. 
$H \left(\pi \left(s_t; \theta \right)\right)$ is used to favor exploration and its strength is controlled by
the factor $\beta$.

The cost function for the estimated value function is:
\vspace{-0.1in}
\begin{equation}
f_v \left(\theta\right)
= \left( R_t - V \left( s_t; \theta \right) \right)^2
\label{eq:costV}
\end{equation}

Additionally, we augment our A3C model to learn current states apart from accumulating rewards for good configurations
using GAE~\cite{schulman2015high}. The deep network, that replaces the transition matrix as the function approximator learns the value of the given state and the policy of the given state. The model uses GAE to compute the value of a given state that not only returns the reward for the model for the given policy decision but also rewards the model for estimating the value of the state. This helps to guide the model to learn the states instead of just maximizing rewards. % moved the # number of parallel agents to design 

\section{Vision}
Our vision is to automate a subset of data center management and operational tasks by leveraging DeepRL. At a high-level, we anticipate the existence of several DeepRL agents, each trained for a specific set of tasks e.g. traffic-engineering, energy-savings, or topology-augmentations. Each agent will run as an application atop an SDN controller. The use of SDN provides the agents with an interface for gathering their required network state and a mechanism for enforcing their actions. 
For example, \sysname should be able to assemble the traffic matrix by polling the different devices within the network, compute a decision for how the optical switch should be configured to best accommodate the current network load, and  reconfigure the network.

\begin{figure}[!tb]
\centering
\includegraphics[width=0.7\linewidth]{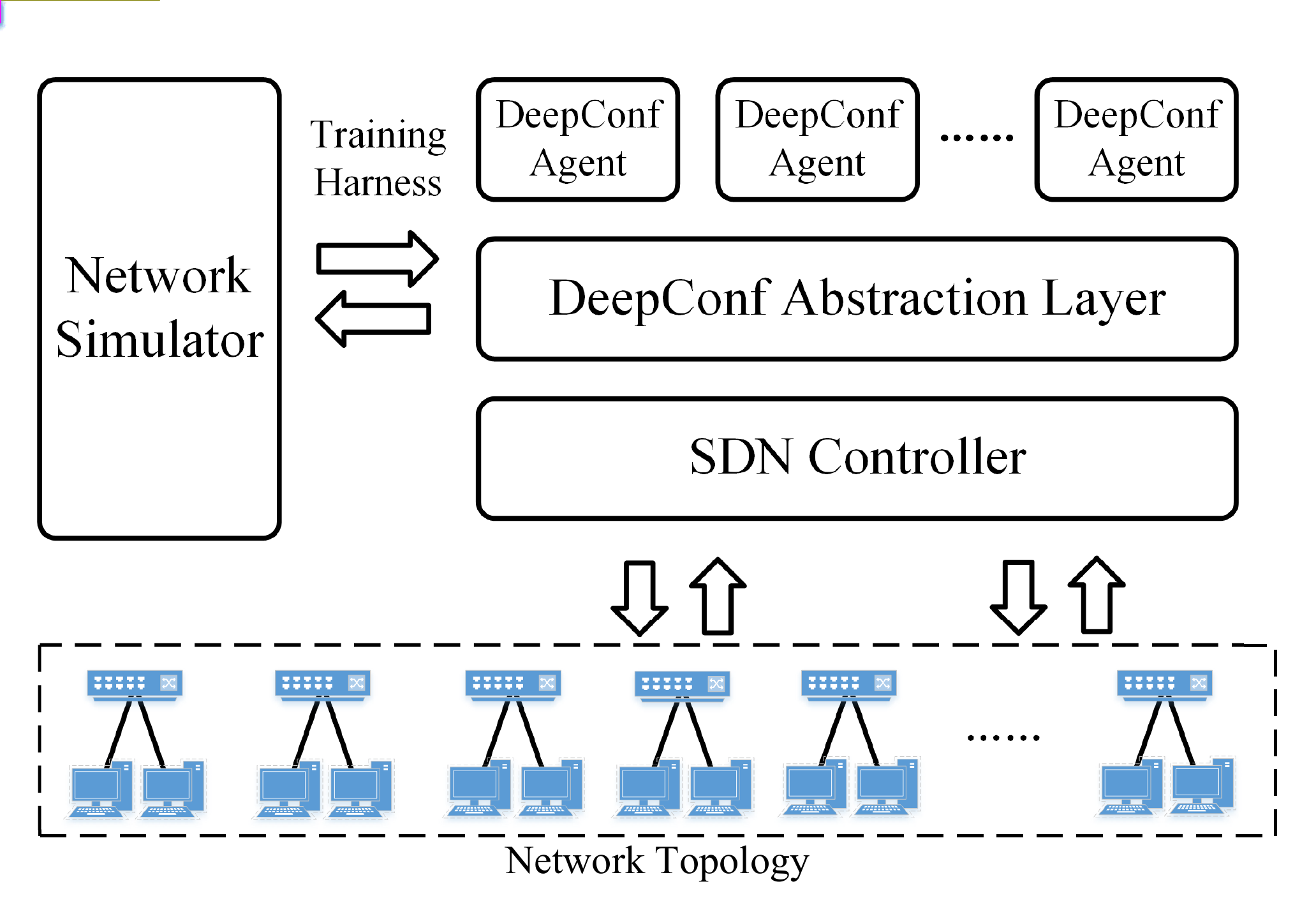}
\vspace{-0.2in}
\caption{\small \bf \sysname Architecture}
\vspace{-0.25in}
\label{f:architecture}
\end{figure}

At a high-level, \sysname's architecture consists of three components (Figure~\ref{f:architecture}): the network simulator to enable offline training of the DeepRL agents, the \sysname abstraction layer to facilitate communication between the DeepRL agents and the network, and the DeepRL agents, called \sysname-agents, which encapsulate data center functionality.

\parab{Applying learning:} The primary challenges in applying machine learning to network problems are (1) the deficiency of training data pertaining to operator and network behavior and (2) the lack of models and loss functions that can accurately model the problem and generalize to unexpected situations. This shortage presents a roadblock for using supervised-based approaches for training ML models. To address this issue, \sysname uses RL where the model is trained by exploring different network states and environments generated by the surplus of simulators available in the network community. Coupled with the wide availability of network job traces, this allows for \sysname to learn a highly generalizable policy.

\parab{\sysname Abstraction Layer:} Today's SDN controllers expose a primitive interface with low-level information. The \sysname applications will instead require high-level models and interfaces. For example, our agents will require interfaces that provide control over paths rather than over flow table entries. While emerging approaches~\cite{sol:nsdi16} argue for similar interfaces, these approaches do not provide a sufficiently rich set of interfaces for the broad range of agents we expect to support and do not provide composition primitives for safely combining the output from the different agents. Moreover, existing composition operators~\cite{pyretic,covisor,frenetic} assume that the different SDN applications (or \sysnameagent in our case) are generating non-conflicting actions -- hence these operators can not tackling conflict actions.  SDN Composition approaches~\cite{oreo,athens:conext14,corybantic} that do tackle conflicting actions, require significant rewrite of the SDNApp which we are unable to do because \sysnameagent are rewritten within the DeepRL paradigm.

More concretely, we require high-layer SDN abstractions that enable to RLAgents to more easily learn and act of the network. Additionally, we require novel composition operators that can reason about and tackle conflicting actions generated by RLAgents. 

\parab{Domain-specific Simulators:} (Low hanging fruit) Existing SDN research leverages a popular emulation platform, Mininet, which fails to scale to large experiments. A key requirement for employing DeepRL is to have efficient and scalable simulators that replays traces and enables learning from these traces. We extend flow-based simulators to model the various dimensions that are required to train our models. To improve efficiency, we explore techniques that partition the simulation and enables reuse of results --- in essence, to enable \textit{incremental simulations}.

In addressing our high-level vision and developing solutions to the above challenges, there are several high-level goals that a production-scale system must address: (1) our techniques must generalize across topologies, traffic matrixes, and a range of operational settings, e.g., link failures; (2) our techniques must be as accurate and efficient as existing state-of-the-art techniques; and (3) our solutions must incur low operational overheads, e.g., minimizing optical switching time or TCAM utilization.

%!TEX root = ../main.tex
\section{Design}
\label{sec:design}
In this section, we provide a broad description of how to define and structure existing data center network management techniques as RL tasks, then describe the methods for training the resulting \sysnameagents.

\subsection{\sysname Agents} 
In defining each new \sysnameagent, there are four main functions that a developer must specify: state space, action space, learning model, and reward function. The action space and reward are both specific to the management task being performed and are, in turn, unique to the agent. Respectively, they express the set of actions an agent can take during each step and the reward for the actions taken. The state space and learning models are more general and can be shared and reused across different \sysnameagents. This is because of the fundamental similarities shared between the data center management problems, and because the agents are specifically designed for data centers. 

In defining a \sysname agent for the topology augmentation problem, we (1) define state-spaces specific to the topology, (2) design a reward function based on application level metrics, and (3) define actions that correlate to activating/de-activating links.

\parab{State Space: } In general, the state space consists of two types of data -- each reflecting the state of the network at a given point in time. 
First, the general network state that all \sysnameagents require: the network's traffic matrix (TM) which contains information on the flows which will executed during the last $t$ seconds of the simulation.

Second, a \sysnameagent specific state-space that captures the impact of the actions on the network. For example, for the topology augmentation problem, this would be the network topology -- note, the actions change the topology. Whereas for the traffic engineering problem, this would be a mapping of flows to paths -- note, the actions change a flow's routes.

\parab{Learning Model:}
Our learning model utilizes a Convolutional Neural Network (CNN) to compute policy decisions. 

\begin{figure}[!tb]
\centering
\includegraphics[width=0.7\columnwidth]{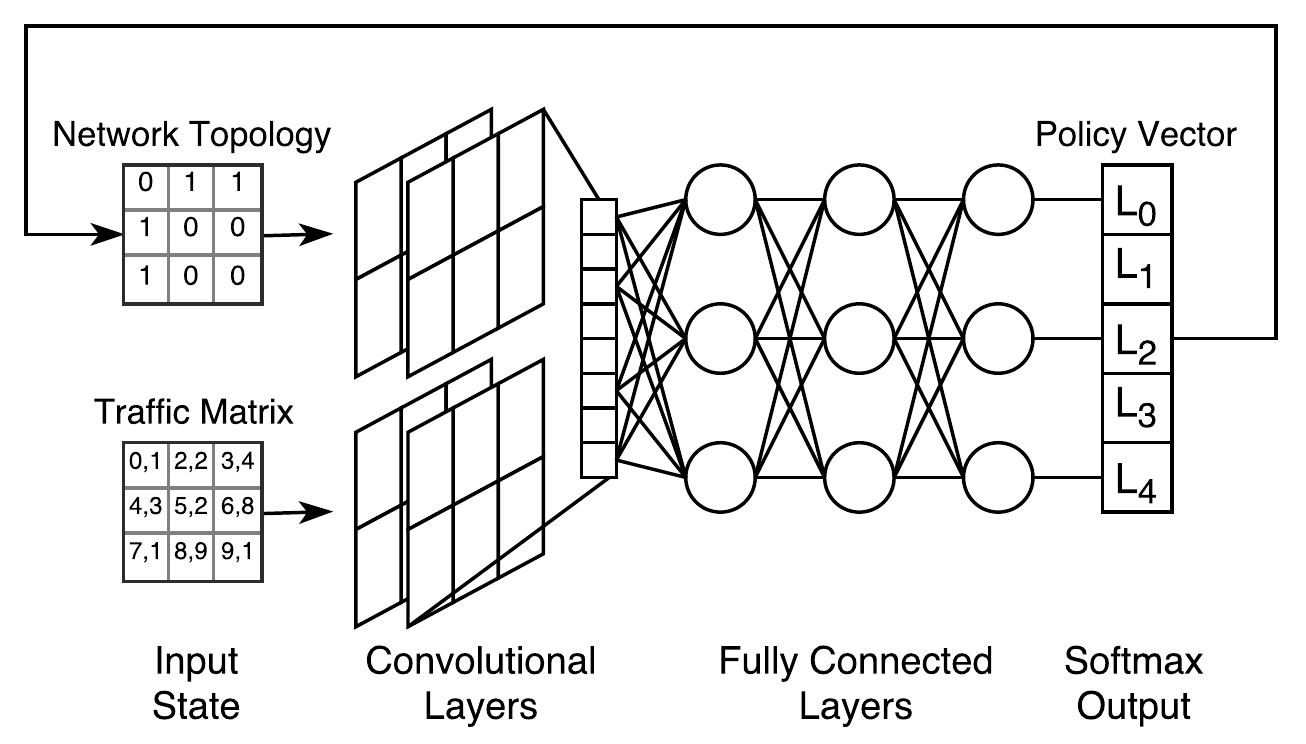}
\vspace{-0.2in}
\caption{\small \bf The CNN model utilized by the RL agent.}
\vspace{-0.1in}
\label{fig:rl_model}
\end{figure}

The exact model for a \sysnameagent depends on the number of state-spaces used as input. In general, the model will have as many CNN-blocks as there are state spaces -- one CNN-block for each state space. The output of these blocks are concatenated together and input into two fully connected layers, followed by a softmax output layer. For example, for the topology-augmentation problem, as observed in Figure~\ref{fig:rl_model}, our \sysnameagent has two CNN blocks to operate on both the topology and the TM states spaces in parallel. This allows for the lower CNN layers to perform feature extraction on the input spatial data, and for the fully connected layers to assemble these features in a meaningful way.

\subsection{Network Training}
\begin{figure}[!tb]
\centering
\includegraphics[width=0.7\linewidth]{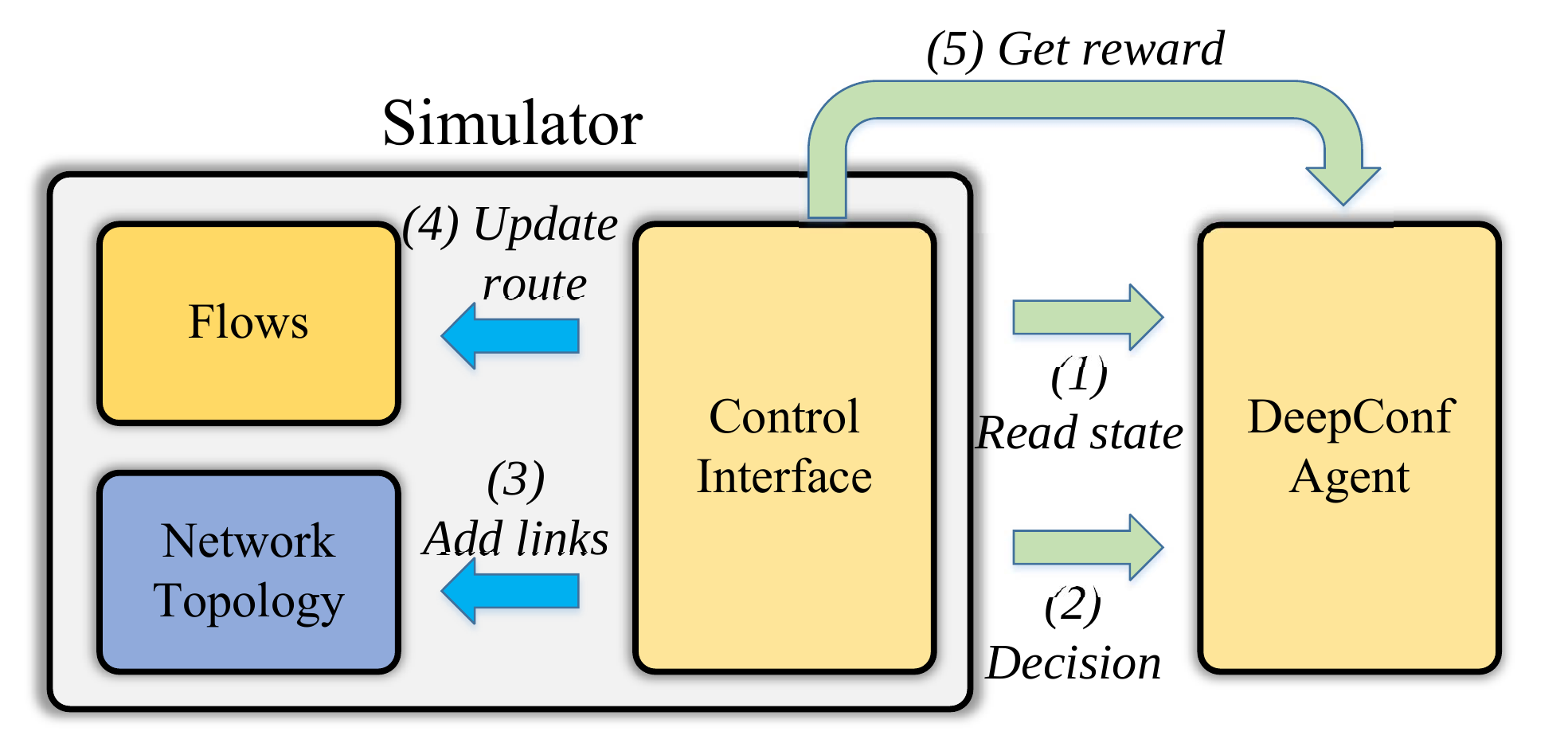}
\vspace{-0.2in}
\caption{\small \bf \sysnameagent model training.}
\vspace{-0.2in}
\label{fig:learning}
\end{figure}

To train a \sysnameagent, we run the agent against a network simulator. The interaction between the simulator (described in Section~\ref{s:eval:setup}) and RL agent can be described as follows (Figure~\ref{fig:learning}): (1) The \sysname agent receives state $s_i$ from the simulator at training step $i$. (2) The \sysnameagent uses the state information to make a policy decision about the network, and returns the selected actions to the simulator. For the \sysnameagent for the topology augmentation problem, called the Augmentation-Agent, the actions are the links to activate and hence a new topology. (3) If the topology changes, the simulator re-computes the paths for the active flows. (4) The simulator executes the flows for $t$ seconds. (5) The simulator returns the reward $r_i$ and state $s_{i+1}$ to the \sysnameagent, and the process restarts.

\textbf{Initialization} During the initial training phase, we force the model to explore the environment by randomizing the selection process --- the probability that an action is picked corresponds to the value of the index which represents the action. For instance, with the Augmentation-Agent, link $i$ has probability $w_i$ of being selected.  As the model becomes more familiar with the states and corresponding values, the model will better formulate its policy decision. At this point, the model will associate a higher probability with the links it believes to have a higher reward, which will cause these links to be selected at a higher frequency. This methodology allows for the model to \textit{reinforce} its decisions about links, while the randomization helps the model avoid local-minima. 

\textbf{Learning Optimization:} To improve the efficiency of learning, the RL agent maintains a log containing the state, policy decision, and corresponding reward. The RL agent performs \textit{experience replay} after $n$ simulation steps. During replay, the log is unrolled to compute the policy loss across the specified number of steps using Equation \ref{eq:costPi}. The agent is trained using Adam stochastic optimization~\cite{DBLP:KingmaB14}, with an initial learning rate of $10^{-4}$ and a learning rate decay factor of 0.95. We found that a smaller learning rate and low decay helped the model better explore the environment and form a more optimal policy.\\

\section{Use Case: Augmentation Agent}

More formally defined, in the topology augmentation problem the data center consists of a fixed hierarchical topology and an optical switch, which connects all the top-of-rack switches. While the optical switch is physically connected to all ToR switches, unfortunately, the optical switch can only support a limited number of active links. Given this limitation, the rules for the topology problem are defined as:
\begin{itemize}
\item The model must select $n$ links to activate at a given step during the simulation. 
\item The model receives a reward based on the link utilization and the flow duration. 
\item The model collects the reward on a per-link basis after $t$ seconds of simulation. 
\item All flows are routed using equal-cost multi-path routing (ECMP). 
\end{itemize}

\noindent\textbf{State Space: }The agent specific state space is the network topology, which is represented by a sparse matrix where entries within the cells correspond to active links within the network. 

\noindent\textbf{Action Space: }The RL agent interacts with the environment by adding links between edge switches. The \textit{action space} for the model, therefore, corresponds to the different possible link combinations and is represented as an $n$ dimensional vector. The values within the vector correspond to a probability distribution, where $w_i$ is equal to the probability of link $i$ being the optimal pick for the given input state $s$. The model selects the highest $n$ values from this distribution as the links that should be added to the network topology.\\

\noindent\textbf{Reward: }The goal of the model can be summarized as: (1) Maximize link utilization and (ii) Minimize the average flow-completion time. 

With this in mind, we formulate our reward function as:
\begin{equation}
R(\Theta, s, i) = \sum_{f\in{F}}\sum_{l\in{f}}\frac{b_f}{t_f}
\end{equation}
Where $F$ represents all active and completed flows during the previous iteration step, $l$ represents the links used by flow $f$, $b_f$ represents the number of bytes transferred during the step time, and $t_f$ represents the total duration of the flow. The purpose of this reward function is to reward for high link utilization but penalize for long lasting flows. The design of this function has the effect of guiding the model towards completing large flows within a smaller period of time.\\

\section{Evaluation}
\label{sec:evaluation}
In this section, we analyze \sysname under realistic workload with representative topologies.

\parab{Experiment Setup}
\label{s:eval:setup}

We evaluate \sysname on a trace driven flow-level simulator using a large scale map-reduce traces from Facebook~\cite{fb:trace}. 

 We evaluate two state-of-the-art clos-style data center topologies: K=4 Fat-tree~\cite{al2008scalable} and VL2~\cite{vl2}. In our analysis, we focus on flow completion time (FCT) a metric which captures the duration between the first and last packet of a flow. We augment both topologies by adding an optical switch with four links.  Here we compare \sysname against the optimal solution derived from a linear program --- Note: this optimal solution can not be solved with larger topologies~\cite{microte:conext11}.

\label{s:eval:model}
\noindent\textbf{Learning Results: } The training results demonstrate that the RL agent learns to optimize its policy decision to increase the total reward received across each episode.

We observed that the loss decreases as training increases, with the largest decrease occurs during the initial training episodes, a result consistent with the learning rate decay factor employed during training.

\begin{figure}
\centering
\includegraphics[width=0.6\linewidth]{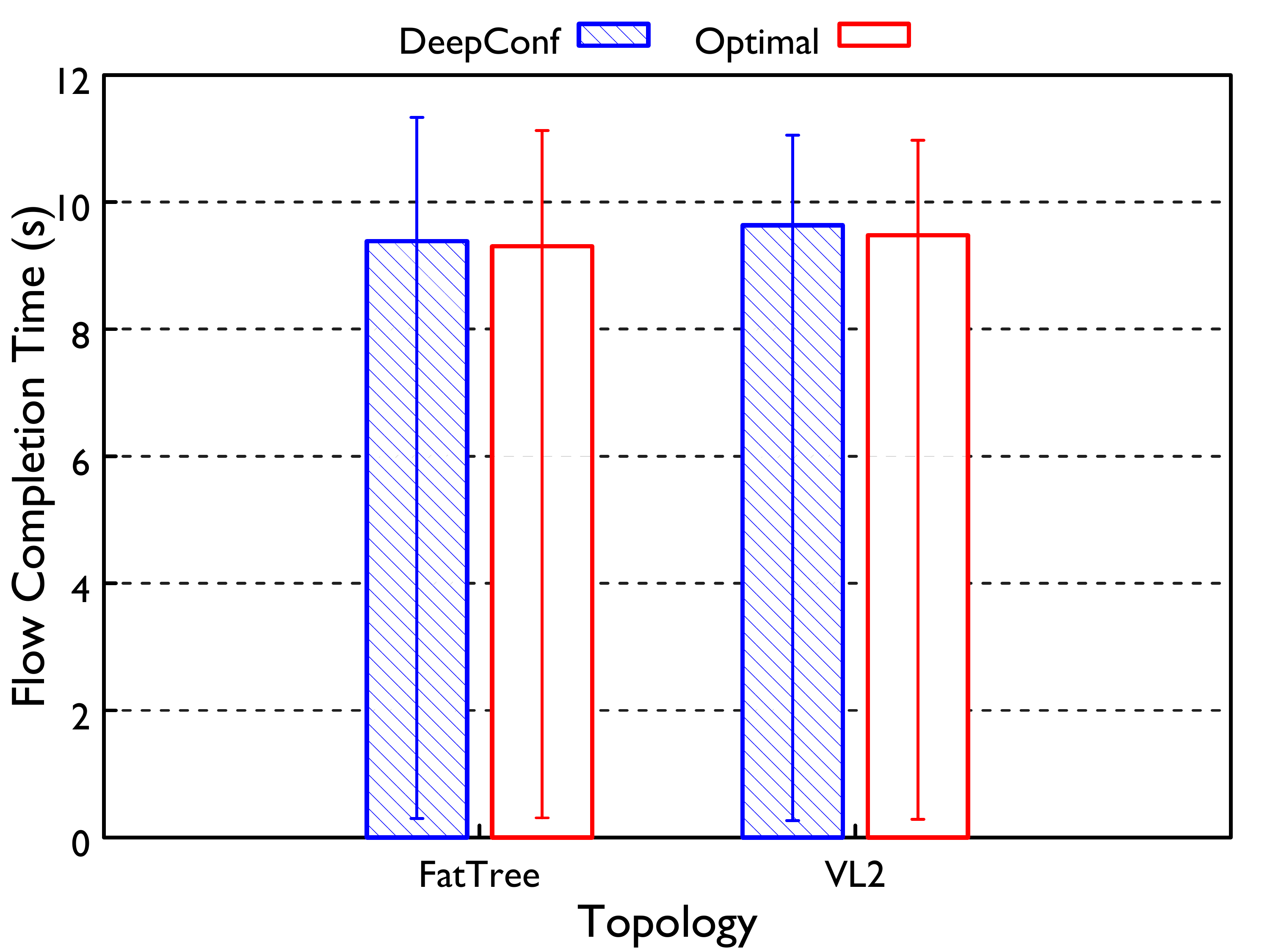}
\vspace{-0.2in}
\caption{Median Flow Completion Time.} \label{fig:fct}
\vspace{-0.25in}
\end{figure}

\noindent\textbf{Performance Results: }
The results, Figure~\ref{fig:fct}, show that \sysname performs comparable with optimal~\cite{wang:sigcomm10,helios:sigcomm10} across representative topologies and workloads. Thus, our system is able to learn a solution that's close to the optimal across a range of topologies. 

\parab{Takeaway} We believe these initial results are promising, and that more work is required in order to understand and improve the performance of \sysname. 

\section{Discussion}
\label{sec:disc}
We now discuss open questions:

\textbf{Learning to generalize:} In order to avoid over-fitting to a specific topology, we train our agent over a large number of simulator configurations. DeepRL agents need to be trained
and evaluated on many different platforms to avoid being overtly specific to few networks and correctly handle unexpected scenarios. Solutions that employ machine learning to address network problems using simulators need to be cognizant of these issues when deciding the training data.

\textbf{Learning new reward functions:} DeepRL methods need appropriate reward functions to ensure that they optimize for the correct goals. For some networks problems
like topology configuration this may be straightforward. However, other problems like routing may require a weighted combination of network
parameters that need to be correctly designed for the agent to operate the network correctly.

\textbf{Learning other data center problems.} In this paper, we focused on problems that center around learning to adjust the topology and routing. Yet, the space of data center problems is much larger.  As part of ongoing work, we are investigating intermediate representations and models for capturing high-level tasks.

\section{Conclusion}
\label{sec:future}
Our high level goal, is to develop ML-based systems that replace existing heuristic-based approach to tackling data center networking challenges. This shift from heuristics to ML, will enable us to design solutions that can adapt to changes in patterns by consumting data and relearning -- an automated task.
In this paper, we take the first steps towards acheving these goals by designing a reinforcement learning based framework, called \sysname, for automatically learning and implementing a range of data center networking techniques.

\small
\bibliography{references}
\bibliographystyle{abbrv} 

\end{document}